\documentclass[10pt]{iopart}  

\usepackage{graphicx}
\usepackage[numbers,sort&compress]{natbib}
\usepackage[colorlinks,citecolor=blue,urlcolor=blue,linkcolor=blue]{hyperref}
\usepackage{booktabs}  
\usepackage{siunitx}
\newcommand{\brac}[1]{\ensuremath{\left( #1 \right)}}  
\newcommand{\gbrack}[1]{\ensuremath{\left\{ #1 \right\}}}  
\newcommand{\avg}[1]{\ensuremath{\left\langle #1 \right\rangle}}  
\newcommand{\abs}[1]{\ensuremath{\left\vert #1 \right\vert}}  

\newcommand{\afz}[1]{``#1''}  
\newcommand{\ER}{Erd\H{o}s-R\'enyi}

\newcommand{\keywords}[1]{\noindent{\it Keywords\/}: #1}

\begin{document}
\title[Energy landscapes for matchings]{Energy landscapes of 
some matching-problem ensembles}
\author{Till Kahlke and Alexander K Hartmann}  
\address{Institut f\"ur Physik, Carl von Ossietzky Universit\"at Oldenburg, 26111 Oldenburg, Germany}
\ead{\mailto{till.kahlke@uol.de}, \mailto{a.hartmann@uni-oldenburg.de}}

\begin{abstract}
The maximum-weight matching problem and the behavior of its energy landscape is numerically investigated. We apply a perturbation method adapted from the analysis of spin glasses. This gives inside into the complexity of the energy landscape of different ensembles.
\ER\ graphs and ring graphs with randomly added edges are considered and 
two types of distributions for the random edge weighs are used.
For maximum-weight matching, fast and scalable algorithms exist,
such that we can study large graphs of more than $10^5$ nodes. Our results 
show that the structure of the energy landscape for standard ensembles of 
matching is simple, comparable to the energy landscape of a 
ferromagnet. Nonetheless, for some of the here 
presented ensembles our results allow for the 
presence of complex energy landscapes in the spirit 
of Replica-Symmetry Breaking.
\end{abstract}


\maketitle

\ioptwocol 

\section{Introduction}
\label{sec:Intro}
Spin glasses are prototypical complex systems widely studied in statistical physics \cite{Binder_1986_spin, Mezard_1987_spin, Fischer_1993_spin, Young_1998_spin, Nishimori_2001_spin, Kawashima_2013_spin}. 
The analytical solution of the Sherrington-Kirkpatrick spin glass model \cite{Sherrington_1975_spin} led to the notion of Replica-Symmetry Breaking (RSB) \cite{Parisi_1979_A, Parisi_1979_fullRSB, Parisi_1980_magnetic, Parisi_1980_sequence, Parisi_1980_order_parameter, Parisi_1983_SpinGlasses}.
Typical for such a complex RSB phase is a rough energy landscape with many local minimal and metastable states divided by high energy barriers \cite{Parisi_1983_SpinGlasses}.
This leads to non-trivial equilibrium and non-equilibrium behavior. 

This concept of complexity is transferable to other problems like neural networks, data analysis or optimization problems \cite{Nishimori_2001_spin, Hartmann_2005_Book, Mezard_2009_information, Moore_2016_computation}.
These problems may show behavior like RSB or other types of complexity 
\cite{Franchini_2022_RSB}.
It has been observed that, e.g., Satisfiability \cite{Kirkpatrick_1994_SAT, Hayes_1997_SAT, Cocco_2001_3SAT}, the traveling salesperson \cite{Gent_1996_TSP, Schawe_2016_TSP, Schawe_2019_RSB}, vertex cover \cite{Hartmann_2000_VC, Hartmann_2001_VC, Weigt_2001_VC, Weigt_2001_VC2} and graph coloring \cite{Mulet_2002_coloring} can feature complex phases for certain ensembles and model parameters. 
These models also exhibit rugged energy landscapes with glassy equilibrium
and non-equilibrium behavior.

These mentioned problems are members of the class of
non-deterministic-polynomial (NP)
problems \cite{Garey_1979_NP}. All known exact  algorithms for
treating these problems exhibit worst-case running times that grow
exponentially with the problem size.  This means, one is rather
limited in the numerical investigations of the equilibrium behavior 
of these problems, because only  comparable small systems can be studied.
This is a pity, because the behavior of these models is very interesting.

On the other hand, in the class of
polynomial (P) problems, the worst case run time growths only
polynomially. Hence, much larger systems can be studied, leading to
numerically more accurate results and better statistics.  Unfortunately,
problems from P
typically feature a simple energy
landscape \cite{Mezard_1987_matching, Middleton_2002_Ising,
Hartmann_2003_2DSpin, Hartmann_2004_2DSpin, Ahrens_2011_Ising,
Muenster_2011_2DIsingSpin}, leading to a rather trivial behavior.
This seems to indicate that complex behavior and convenient numerical
access to the equilibrium behavior do not come together. But it
should be noted that most studies so far where performed for model ensembles,
where the randomness is introduced in a simple and uncorrelated way.
Interestingly, recently  a complex behavior was found for ensembles
which utilize correlations for the model of directed polymers in random media.
For this model a polynomial exact sampling algorithm exists and thus large
systems of up to $10^9$ lattice sites could be studied and strong indications
for replica-symmetry breaking have been observed \cite{Hartmann_2022_polymers}.
Furthermore, for the polynomially tractable problem of increasing subsequences, also called
Ulam's problem, numerical evidence for RSB was found \cite{is_rsb2023}.
Finally, it should be noted that also the XOR-SAT problem, which can be solved
efficiently by Gaussian elimination \cite{ricci-tersenghi2001,mezard2003}, 
exhibits one-step RSB, but this becomes irrelevant in
equilibrium in the thermodynamic limit \cite{ricci-tersenghi2010}.

Thus, by the choice of a suitable ensemble, it seems to be possible
to study some systems with non-trivial behavior in a numerically 
efficient way. In the present work, 
in the spirit of \cite{Hartmann_2022_polymers,is_rsb2023}, we study 
another problem from the class P to seek for indications for complex 
behavior. In particular, we consider the maximum-weight graph-matching problem. 
This model was one of the first optimization problems to be studied from the viewpoint of statistical mechanics \cite{Mezard_1985_matching}. Analyses were extended to arbitrary graphs \cite{Mezard_1987_matching}, finite-size effects \cite{Parisi_2002_matching} and Euclidean edge weights \cite{Mezard_1988_euclidean, Houdayer_1998_euclidean, Caracciolo_2014_euclidean}.

The interests of graph-matching problems in physics also comes from the fact that other problems can be solved by a mapping to suitable matching problems. This has been done for, e.g., 2D spin glasses \cite{Hartmann_2004_2DSpin}, dimer covering \cite{Kenyon_2009_lectures}, negative-weight percolation \cite{Melchert_2014_NWP} and controllability of dynamic networks \cite{Liu_2011_controllability, Menichetti_2014_controllability}.

The remaining of this paper is organized as follows:
In \sref{sec:Model}, graph matching and random graph ensembles used in
this work are introduced. These ensembles are \ER\ graphs and ring
graphs with additional random edges, both with Gaussian distributed or
constant edge weights with uniform noise.
\Sref{sec:Methods} presents the details of the perturbation technique. 
It it explained how one can investigate whether a complex energy landscape is  
present by observing 
the differences between matchings obtained from the original
and slightly perturbed graphs, respectively. We also address the
question whether the perturbation approach can be used to sample
matching statistically correctly. 
The results of the analyses are given in \sref{sec:Results}, where
we start with the results obtained from the perturbation technique.
We summarize and draw final conclusions in \sref{sec:Conclusion}.

\section{Model}
\label{sec:Model}

\subsection{Matching}
An undirected graph $G = \brac{V, E}$ is given by a set of nodes
$i \in V$ and a set of edges $e \in E \subseteq V^{(2)}$, where $n
= \abs{V}$ and $m = \abs{E}$.  Two nodes $i$ and $j$ are
called \emph{adjacent} if $e = \gbrack{i,\,j}$ is a member of $E$. This
edge $e$ called \emph{incident} to $i$ and $j$.  A matching $M$ is a
subset of $E$ in which no two edges are incident to the same node. For
each edge in $M$, its adjacent nodes are called \emph{matched}. Nodes
which are not matched are called \emph{free}. 
A maximum cardinality matching is a
matching with maximizes the number of matched nodes $\abs{M}$.  On a
graph $G$ with edge weights $w(e)$, the weight of the matching
is $W = \sum_{e \in M} w(e)$. A
 maximum-weight matching is a
matching $M$ that maximizes the weight $W$ 
over all possible matchings on $G$. In \sref{sec:Methods} we state
the algorithm we have used to obtain the maximum-weight matching of a given
graph.

\subsection{Random graphs \label{sec:graphs}}
We consider two types of random graph ensembles. Details on the chosen values of the model parameters are given in \sref{sec:Results}.
The first ensemble consists of \ER\ random graphs \cite{Erdos_1959_random}. 
To generate such graphs, one starts with on empty graph of $n$ nodes 
$V=\{1,2,\cdots, n\}$.
Then, for each possible pair of distinct nodes, an edge is created independently 
with probability $p = c/n$, where $c$ is the \emph{connectivity}. The resulting degree distribution is Poissonian.

The second ensemble consist of ring graphs with additional random edges. First, a regular ring graph with mean degree of two is created. This means, the edge
$E$ set initially contains 
the edges $\{i,i+1\}$ for $i=1,\ldots,n-1$ plus the edge $\{n,0\}$.
Next, two nodes $i, j$ are chosen randomly with uniform probability
 and the edge $\gbrack{i, j}$ is created if $i \neq j$ and $\gbrack{i, j}$ does not yet exist in $E$. This process is repeated until $m^+$ edges are added to the graph this way. The number of edges in $G$ is then given by $m = n + m^+$.
Below we will consider the cases $m^+=$ const, $m^+ \in O(\log n)$ and $m^+ \in O(n)$.

Two types of edge weights $w(e)$ are investigated. First,
we consider $w(e) \sim \mathcal{N}(1, 0.01)$, i.e.,
Gaussian distributed random weights with a mean of 1 and a variance of $0.01$.
Second, we study constant weights with additional uniform noise, 
$w(e) = 1 + \varepsilon$. To choose the strength of the additional noise
$\varepsilon$, we mention that
with $w(e) = 1$ for all edges, a maximum-weight matching is equivalent to a 
maximum cardinality matching, since $W = \abs{M}$ holds. 
The purpose of the additive noise $\varepsilon$ is to make the 
optimal solution unique. But beside that, it should not influence 
the size of the optimum matching even for large values of $n$. To ensures 
that, 
$\varepsilon$ is drawn from a uniform distribution where the values 
scale with $n^{-\frac{1}{2}}$, i.e., 
$\varepsilon \sim U(-0.01 \, n^{-\frac{1}{2}},\ 0.01 \, n^{-\frac{1}{2}})$.

\section{Methods}
\label{sec:Methods}
To find a maximum weight matching we use Edmond's Blossom Shrinking algorithm \cite{Cook_1998}, implemented in the LEMON-library \cite{LEMON}.
The algorithm has a worst-case running time $O(n^2m)$, i.e., polynomial,
such that correspondingly large graphs can be treated.

\subsection{Perturbation technique}
\label{sec:perturbation_technique}

When studying the complexity of energy landscapes, a common technique is to apply weak perturbations on the system. To our knowledge, this technique was first used to study spin glasses (see, e.g., \cite{Palassini_2000_spinglass}). Since then, it was adapted for, e.g., random-field Ising models \cite{Zumsande_2008_excitations, Zumsande_2009_excitations} and the 
traveling salesperson problem \cite{Schawe_2019_RSB}.

The basic idea is to first calculate a ground state and the perturb the
system slightly such that by a recomputation a new ground state
is obtained, which is an excited state with respect to the original system.
For the matching problem, we use \emph{edge flips} as a perturbation technique. First, a maximum-weight matching $M_0$ with total weight $W_0$ is calculated. Then, one randomly chosen edge $e$ will be \afz{flipped}. If $e$ is in $M_0$, it is not allowed to be in a matching $M_1$ of the perturbed graph. On the other hand, if $e$ is not in $M_0$, it is enforced to be in $M_1$. In practice, this can be achieved by setting the weight of $e$ to a very small value, e.g., $-2 W_0$, or a very large value, e.g., $2 W_0$, respectively. After $M_1$ is found, the weight of the edge 
will be reset to its
original value such that the weight $W_1$ of $M_1$ is calculated with the
original weights.
In the following, $M_0$ will always denote an optimal matching with
respect to the original edge weights, while $M_i$ with $i > 0$ denote 
 independent matchings resulting from perturbations by 
a single edge flip with respect to $M_0$. $W_0$ and $W_i$ denote the
corresponding weights of the matchings.

A perturbation is weak if the relative difference of $W_0$ and $W_1$ behaves as
\begin{equation}
  \label{eq:quasi_optimal}
  \frac{W_0 - W_1}{W_0} = \Or\brac{\frac{1}{n}}.
\end{equation}
If \eref{eq:quasi_optimal} holds, $M_1$ is \emph{quasi optimal}. Note
that this means that when considering the system in the canonical ensemble, 
i.e., according to the Boltzmann distribution, the matchings $M_0$ and $M_1$
will contribute with the same weight in the thermodynamic limit.

To compare two matchings $M_i$ and $M_j$, 
we apply a similarity measure, also called overlap.
Since matchings are sets, it is convenient to use the
Jaccard index \cite{Jaccard_1912_index}, given by
\begin{equation}
  \label{eq:overlap}
  q_{ij} = \frac{\abs{M_i \cap M_j}}{\abs{M_i \cup M_j}}.
\end{equation}
The distance between two matchings 
is defined as $d_{ij} = 1 - q_{ij}$ \cite{Levandowsky_1971_distance}.
Overlap and distance between optimal matching $M_0$ and matching $M_i$ 
obtained from a perturbation are 
denoted as $q_0$ and $d_0$, respectively, where the second index $i$ is omitted.

A necessary condition for a complex energy landscape is that there exist quasi optimal matchings 
with $d_0 > 0$ in the thermodynamic limit of infinite large graphs \cite{Mezard_1986_replica}.
Hence, we measured $\avg{(W_0 - W_i)/W_0}$ and $\avg{d_0}$ for different 
values of $n$, averaged over random graph realizations. If the 
matchings for the perturbed graphs are quasi optimal and $\avg{d_0}$ converges 
towards zero for increasing values of $n$, a simple energy landscape 
can be assumed. But if $\avg{d_0}$ remains finite, a complex energy 
landscape can not be excluded.

\subsection{Sampling bias test}
\label{sec:sampling_bias_test}

To further investigate whether the matching problem on a given
graph ensemble indeed exhibits a complex energy
landscape, it would be desirable to efficiently 
sample matchings. For correct statistics,
matchings with equal total weight $W$ should be sampled with
equal probability.  
Hence, a sampling 
approach needs to
know the number of distinct matchings for the full problem and
for encountered subproblems.
Unfortunately, the matching problem is in the class
of \#P-complete problems 
\cite{matching_is_numberP,vazirani_2002_numberP}. This means,
with all known algorithms, it
requires to enumerate all matchings, which are typically exponentially
many. Hence, such an approach is computationally demanding.

We can, in principle, use matchings obtained by the 
perturbation method as samples.
Unfortunately, edge flips create an unknown bias to the sampling 
because various configurations are prohibited, which depends on the flipped
edge.
It could nonetheless be the case that bias errors average out, 
especially for large graphs.
To assess this, we have performed the following investigation.

For a given graph, here we study \ER\ graphs, 
a set of edge weights of constant weights with noise 
$w = 1 + \varepsilon$ is drawn, see \sref{sec:graphs}. Then, an 
optimal matching according to this set of weights is calculated. Next, 
multiple times
a perturbed graph is obtained, using random edge flips, and again
matchings are calculated. This process is then repeated for multiple 
realizations of the initial noise.

The obtained matchings are collected in a sample histogram. Here, we
include only matchings with the same cardinality as the
optimal  matching. The resulting histogram is then used
to compare the relative frequencies $Q(M)$ to the case where all
matchings would have been drawn with equal probability. That is $P(M)
= 1/n_\mathrm{M}$, where $n_\mathrm{M}$ is the number distinct
 matchings.  As the number of
matchings grows exponentially with the number of edges, this procedure
is only appropriate for small graphs.

To compare $P$ and $Q$ quantitatively, two distance measures are used. The first one is the \emph{Kullback-Leibler divergence} $D_\mathrm{KL}$ defined as
\begin{equation}
  D_\mathrm{KL}(P,Q) = \sum_{i=1}^{n_\mathrm{M}} P(M_i) \ln\brac{\frac{P(M_i)}{Q(M_i)}}.
\end{equation}
The second measure is the mean relative error $\mathrm{MRE}$ given by
\begin{eqnarray}
  \mathrm{MRE}(P,Q) &= \sum_{i=1}^{n_\mathrm{M}} P(M_i) \frac{\sqrt{\brac{P(M_i) - Q(M_i)}^2}}{P(M_i)} \\
  &= \sum_{i=1}^{n_\mathrm{M}} \sqrt{\brac{P(M_i) - Q(M_i)}^2},
\end{eqnarray}
which is the averaged relative error per matching.

If $P$ and $Q$ become more equal for larger graphs, the values of $D_\mathrm{KL}$ and $\mathrm{MRE}$ will become smaller. In particular, if the sampling
approaches the uniform one, the distance measures should converge to zero.
If this was the case,  the
overlap distributions $P(q)$  could be investigated in a meaningful way.
In case the distribution of overlaps remains broad in the
limit of large graphs, this would indicate that a complex behavior 
is significant in equilibrium.
On the other hand, if the values of $D_\mathrm{KL}$ and $\mathrm{MRE}$ 
increase with larger graphs, the sampling bias errors can not be neglected.
In this case, a limiting nonzero value of the distance  $d_0$
would only indicate that significantly different matchings exists,
but one would not know by using the present approach 
whether their weights are strong enough
to influence the equilibrium behavior.

\section{Results}
\label{sec:Results}
The following results are obtained by performing 
simulations \cite{Alexander_2015_practical_guide} for the two graph ensembles, various values of parameters and graph sizes in the range of $n = 7$ to $\num{131072}$. If not otherwise specified, all results are averaged over 100 different random graph realizations and 100 different edge-flip perturbations
 and resulting 
 matchings for each graph realization.
Partially, the {\tt GNU Parallel} tool was used to distribute the 
simulations over several CPUs \cite{parallel}. All data fits were
 performed as recommended in Ref.~\cite{Young_2014_fits}.

\subsection{Perturbation}
\label{sec:Perturbation}

\begin{figure}[htb]
  \includegraphics{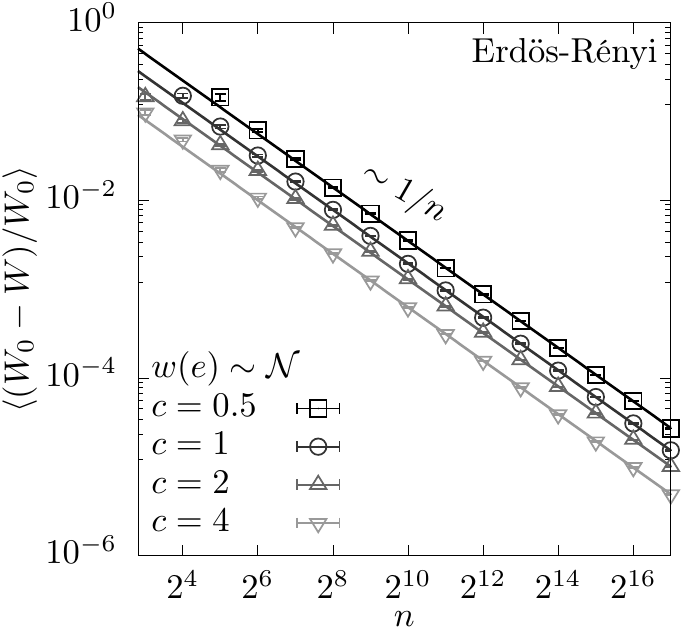}
  \caption{Verification that the matchings on perturbed graphs are quasi optimal, as $\avg{(W_0-W)/W_0}$ as a function of $n$ behaves as $\Or(1/n)$. Shown here are results for \ER\ graphs with Gaussian distributed random weights and connectivities $c \in \gbrack{0.5,\, 1,\, 2,\, 4}$. The straight lines represent fits in the form of \eref{eq:fit_quasi_optimal}.}
  \label{fig:ER_w0_w_over_n_Gauss}
\end{figure}

We first verify that the matchings of the perturbed graphs are quasi optimal.
Our results show that for both types of random graphs, \ER\ and ring graphs, as well as for both types of edge weights, the matchings for the 
perturbed graphs are quasi optimal as \eref{eq:quasi_optimal} holds.
In \fref{fig:ER_w0_w_over_n_Gauss},  
results are shown for \ER\ graphs with connectivities $c \in \gbrack{0.5,\, 1,\, 2,\, 4}$. Included are results from fits using
\begin{equation}
  \label{eq:fit_quasi_optimal}
  \avg{\brac{W_0 - W_i}/W_0} = \alpha \, \frac{1}{n}.
\end{equation}
 In the other studied cases the results look similar, hence they
are not shown here. Note that for ring graphs, the data deviates from the fit 
for small sizes $n \leq 2^8$. Since \eref{eq:quasi_optimal} is required 
to hold only asymptotically, the condition for quasi optimality is fulfilled
for this case as well. 

\begin{figure}[tb]
  \includegraphics{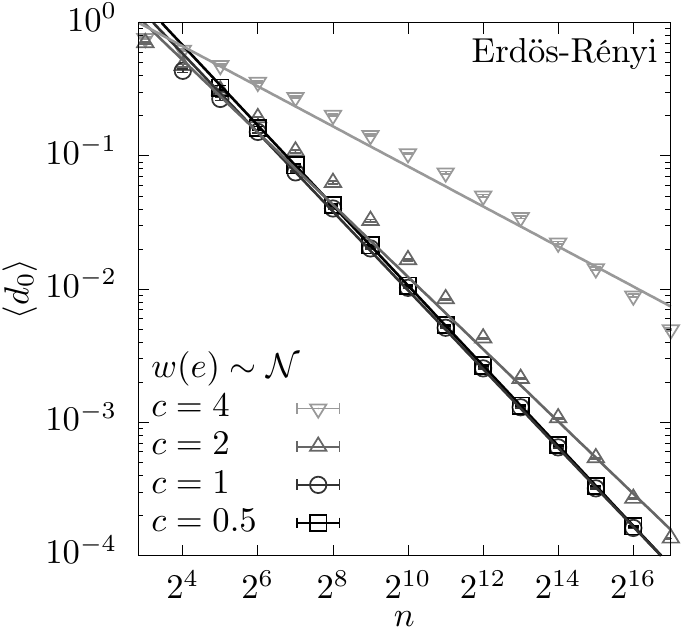}
  \caption{Averaged distance $\avg{d_0}$ 
between matchings on perturbed graphs and optimal matchings as a function of $n$ for \ER\ graphs with Gaussian distributed random weights and connectivities $c \in \gbrack{0.5,\, 1,\, 2,\, 4}$. The straight lines represent power low fits in the form of \eref{eq:fit_d0}.}
  \label{fig:ER_d0_over_n_Gauss}
\end{figure}

We next look at the behavior of $d_0$ and start with the case of 
Gaussian distributed
weights.  
The results for \ER\ graphs are shown
in \fref{fig:ER_d0_over_n_Gauss}.
One observes that 
the distance $\avg{d_0}$ approaches  zero
for increasing $n$.
To confirm this behavior
systematically, a power law fit with offset $d_\infty$ is used,
\begin{equation}
  \label{eq:fit_d0}
  \avg{d_0}(n) = d_\infty + \beta_1 \, n^{\beta_2}\,,
\end{equation}
resulting in good fits. The resulting fit parameters for all studied cases 
are reported in \tref{tab:fitparameters_d0}.

\begin{table}[tb]
  \caption{Values of the fit parameters for the fits in the form of $\eref{eq:fit_d0}$ to determine $d_\infty$. Fixed values are marked with *. Results are given for various values of parameters (first column) and edge weights (second column) for \ER\ graphs and ring graphs with even or odd number of edges. A value of $d_\infty = 0$ corresponds to a simple energy landscape. But for $d_\infty > 0$, a complex behavior can not be excluded. Values for $\beta_1$ and $\beta_2$ are given for completeness.}
  \label{tab:fitparameters_d0}
  \centering
  \begin{tabular}{S[table-format=1.1]cS[table-format=1.4(1),table-text-alignment=left]S[table-format=1.4(1),table-text-alignment=left]S[table-format=1.4(1),table-text-alignment=left]}
  \toprule
    \multicolumn{5}{c}{\ER\ graphs} \\
  \toprule
      {$c$} & {$w$} & {$d_\infty$} & {$\beta_1$} & {$\beta_2$} \\
  \midrule
      0.5 & $\sim \mathcal{N}$ & {0*} & 10.7(1) & 0.997(1) \\
      1   & $\sim \mathcal{N}$ & {0*} & 8.8(1)  & 0.982(1) \\
      2   & $\sim \mathcal{N}$ & {0*} & 6.27(7) & 0.898(1) \\
      4   & $\sim \mathcal{N}$ & {0*} & 2.18(2) & 0.446(1) \\
    \midrule
      0.5 & $1 + \varepsilon$ & {0*} & 10.8(2) & 0.998(2) \\
      1   & $1 + \varepsilon$ & {0*} & 8.9(1)  & 0.983(1) \\
      2   & $1 + \varepsilon$ & {0*} & 6.36(7) & 0.901(1) \\
      4   & $1 + \varepsilon$ & {0*} & 2.65(2) & 0.499(1) \\
    \bottomrule
      \multicolumn{5}{c}{ring graphs, $n$ even}  \\
    \toprule
      {$m^+(n)$} & {$w$} & {$d_\infty$} & {$\beta_1$} & {$\beta_2$} \\
    \midrule
      0  & $\sim \mathcal{N}$ & {0*} & 83(3)  & 0.989(3)    \\
      1  & $\sim \mathcal{N}$ & {0*} & 81(3)  & 0.986(3)    \\
      10 & $\sim \mathcal{N}$ & {0*} & 80(2)  & 0.985(3)    \\
    \midrule
      0  & $1 + \varepsilon$ & {1*}     & {0*}   & {0*}       \\
      1  & $1 + \varepsilon$ & 0.947(2) & {0*}   & {0*}       \\
      10 & $1 + \varepsilon$ & 0.583(3) & 2.5(5) & 1.16(9)    \\
      {$2\,\ln(n)$} & $1 + \varepsilon$ & 0.345(5) & 1.00(3)  & 0.35(1)   \\
      {$4\,\ln(n)$} & $1 + \varepsilon$ & 0.255(3) & 1.31(3)  & 0.42(1)   \\
    \bottomrule
      \multicolumn{5}{c}{ring graphs, $n$ odd}  \\
    \toprule
      {$m^+(n)$} & {$w$} & {$d_\infty$} & {$\beta_1$} & {$\beta_2$} \\
    \midrule
      0  & $\sim \mathcal{N}$ & {0*} & 81(3)  & 0.987(3)    \\
      1  & $\sim \mathcal{N}$ & {0*} & 79(3)  & 0.984(3)    \\
      10 & $\sim \mathcal{N}$ & {0*} & 84(2)  & 0.991(3)    \\
    \midrule
      0  & $1 + \varepsilon$ & 0.544(3) & 1.5(5) & 1.1(2)  \\
      1  & $1 + \varepsilon$ & 0.523(3) & 2.0(5) & 1.1(1)  \\
      10 & $1 + \varepsilon$ & 0.397(3) & 2.2(2) & 0.92(4) \\
      {$2\,\ln(n)$} & $1 + \varepsilon$ & 0.295(3) & 1.42(6)  & 0.56(2)   \\
      {$4\,\ln(n)$} & $1 + \varepsilon$ & 0.222(3) & 1.50(4)  & 0.51(1)   \\
  \bottomrule
  \end{tabular}
\end{table}

For Gaussian distributed weights, the value of $d_\infty$ is set to a fixed value of zero in all cases to obtain reasonable results. Fits with a freely estimated $d_\infty$ result in negative values for this parameter, i.e., nonphysical results.
The observed behavior shows that the matchings 
for the perturbed graphs share 
most of their edges with the matching for the original graphs,
increasingly with growing system size.
This can be explained by the application of a 
non-uniform weight distribution. To maximize $W_0$, the matching algorithm focuses on few edges with large weights. Now for a matching with perturbed weights, if one edge is flipped, a large total weight is still achieved by using most of the same edges with large weights as in $M_0$. Consequently, only a few edges change compared to $M_0$ to compensate the edge flip. This results in a small difference between optimal matching and matching of the perturbed graph that vanishes in the limit of infinity large graphs.

Overall, the results for Gaussian weights indicate
 a simple structure of the energy
landscape.  The results for ring graphs are 
similar and therefore not shown here.  
As constant values of $m^+$ result in $d_\infty = 0$, we omitted to 
investigate values of $m^+ \in O(\log n)$ or $m^+ \in O(n)$ for Gaussian distributed weights.

We next look at constant edge weights with additional noise.
Since here all edge weights are more similar to each other, the existence
of very different matchings with very similar weight, i.e., a more
complex energy landscape, can be anticipated.
For \ER\ graphs the results for $\avg{d_0}$ are similar to the above discussed
 case with Gaussian distributed weights, and therefore no shown here.
Again, fixed values of $d_\infty = 0$ are needed to obtain physically 
meaningful results for the fits according to (\ref{eq:fit_d0}).
Thus, also for this case no sign of a complex energy landscape can be found. 

The argumentation for Gaussian weights no longer holds here since each edge is almost equally important for the total weight. But here, the topological constrains of \ER\ graphs explain the behavior: Since loops are of order $\log(n)$
for \ER\ graphs, local changes of the matching do not spread far, resulting
only in small overall differences between the optimum matching for the
original and the perturbed graph. 
Hence, the difference $\avg{d_0}$ vanishes for $n \to \infty$.

\begin{figure}[tb]
  \includegraphics{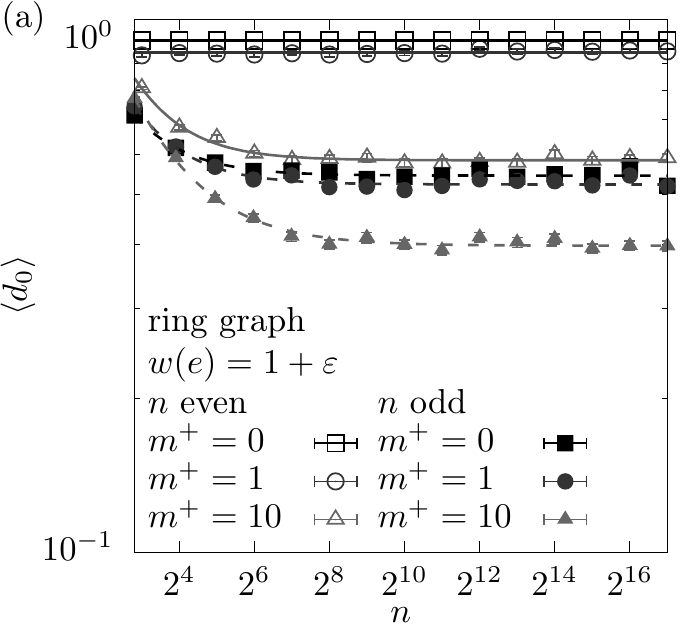}
  \includegraphics{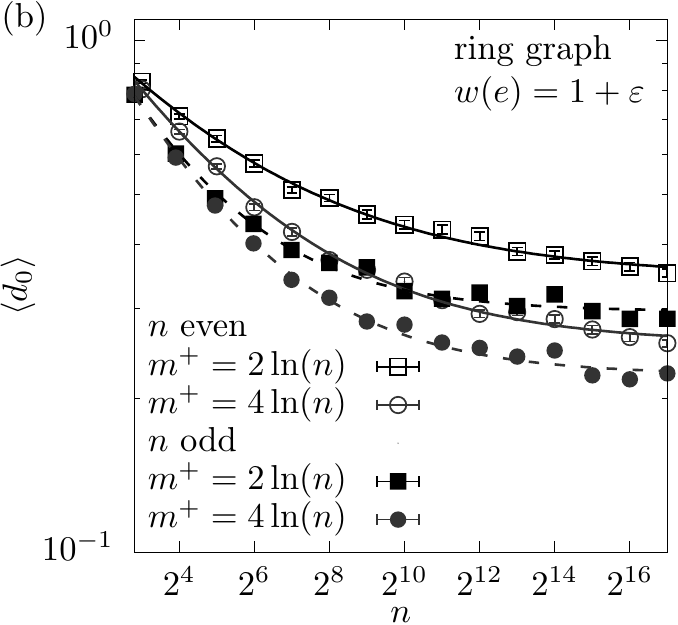}
  \caption{Averaged distance $\avg{d_0}$ between matchings on perturbed graphs and optimal matchings  as a function of $n$ for ring graphs. (a) shows results for constant edge weights with additional noise and values of $m^+ \in \gbrack{0,\, 1,\, 10}$ and (b) shows results for $m^+ = l^+ \ln(n)$ with $l^+ \in \gbrack{2,\, 4}$. Blank symbols indicate even number of nodes $n$, while filled symbols are used for odd values of $n$. The straight or dashed lines represent power low fits with offset $d_\infty$ in the form of \eref{eq:fit_d0}, respectively.}
  \label{fig:Ring_d0_over_n_noise}
\end{figure}

On the other hand, for ring graphs the behavior is different, 
as it  can also be seen in \fref{fig:Ring_d0_over_n_noise}.
Here, a value of $d_\infty > 0$ was needed for a good fit in all studied cases. 

To understand this behavior, consider ring graphs with an even number 
$n$ of nodes and no random edges, i.e., $m^+ = 0$. Since matchings along
paths appear as alternations of matched and free edges, 
the simple ring graph structure enforces  that all matchings $M_i$ with $i > 0$ are given 
by $E \setminus M_0$, i.e., all matched edges are replaced by
free ones, and vice-versa.  Hence, trivially  $d_0 = 1$ holds.

If $n$ is odd, one node remains free and there are $n$ maximum cardinality matchings. $M_0$ is given by the maximum cardinality matchings with the largest total weight. Using $E \setminus M_0$ for a perturbed graph would results in lower cardinality of the matchings and hence lower total weight. However, $n-1$ other matchings remain to compensate for the edge flip, each of which leaves another node free. Thus, one average about half of the edges change in the optimum matching
when the graph is perturbed. This can be compared to a simple spin system:
for a ferromagnet when anti-periodic boundary conditions are introduced,
two domain walls will be created, one where the new boundary conditions are
located, 
the other one anywhere in the system. This also results in many
different configurations, but the system still actually does not behave 
in a complex way.
In conclusion, that $d_\infty$ stays finite for $m^+ = 0$ for the
simple ring graph is not 
resulting from a complex energy landscape, but rather from the 
simple structure of the graphs.

When adding random edges, the situation becomes more interesting. The perturbation leads for the cases of $m^+ =$ const and $m^+ \in O(\ln n)$ 
to various matchings and more complex overlaps between the matchings become possible. The finite value of $d_\infty$ can no longer be explained by the graph structure alone. Hence, we observe quasi optimal solutions that differ from the optimal solution by an extensive number of edges. This is a necessary condition for complexity \cite{Mezard_1986_replica}. Thus a complex energy landscape 
may exist for this ensemble.

We also performed simulations with ring graphs where the number of random 
edges $m^+$ grows linearly with $n$. Here, the ring graphs become increasingly equal to \ER\ graphs for large values of $n$ and the ring structure becomes less important. Hence, similar findings as for \ER\ graphs were obtained, i.e.,
we observed $\langle d_0 \rangle \to 0$ for $n\to \infty$. Therefore,
we do not report further results for this case.

\subsection{Sampling}
\label{sec:Sampling}

Our previous results show that there exists ensembles for the matching
problem, where optimum matchings exists, which differ by $O(n)$ variables
but almost have the same energy, i.e., are quasi optimal. This is
a necessary condition for the existence of a 
complex energy landscape. As mentioned above,
it would be desirable to sample these matchings and compare them
to each other, to understand the energy landscape better, in particular
in the thermodynamic limit. For this purpose sampling with equal 
probability, i.e., in an unbiased way, is needed.

For the sampling bias test, \ER\ graphs with $1+\epsilon$ weights, 
$c = 2$ and $n$ between 4 and 32 are used. Note that 
fully enumerating all solutions and sampling with good statistics 
is only feasible for such small sizes. 
For each graph \num{1000} different sets of edge weights are used and 
\num{1000} matchings for the perturbed graphs are calculated for each of 
these weight sets, resulting in $10^6$ matchings for each graph.

\begin{figure}[tb]
  \includegraphics{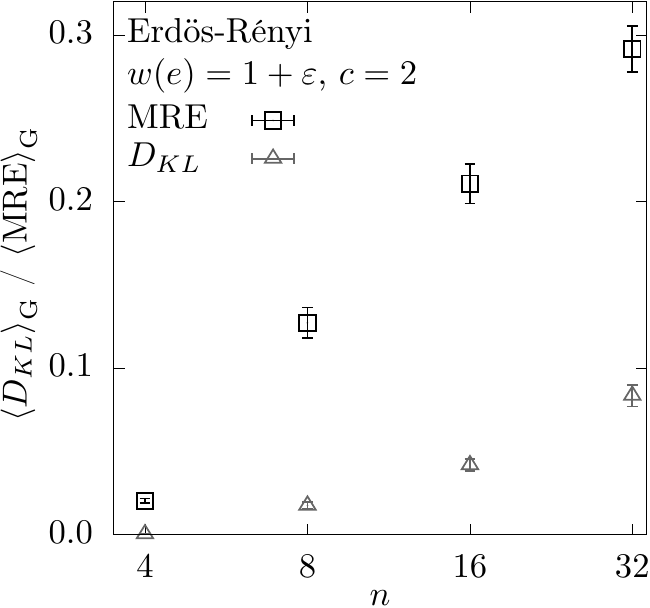}
  \caption{Result of the sampling bias test. Averaged Kullback-Leibler divergence $\avg{D_\mathrm{KL}}_\mathrm{G}$ and mean relative error $\avg{\mathrm{MRE}}_\mathrm{G}$ both systematically increase with larger values of $n$. The results are obtained from \ER\ graphs with a connectivity of $c = 2$ and constant edge weights with additional noise.}
  \label{fig:sampling_check}
\end{figure}

The frequencies $Q(M_i)$ of occurrences of these matchings are measured
and compared to the required uniform distribution as
described in section \ref{sec:sampling_bias_test}.
The obtained values of $D_\mathrm{KL}$ and $\mathrm{MSE}$ are averaged over 100 different random graph realizations. This averaging is denoted by $\avg{.}_\mathrm{G}$. Results are shown in \fref{fig:sampling_check}. Here it can be seen that both $\avg{D_\mathrm{KL}}_\mathrm{G}$ and $\avg{\mathrm{MSE}}_\mathrm{G}$ increase for increasing values of $n$. Hence, the sampling bias does not average out. It can also be seen that the error bars increase with larger values of $n$.
This is the case, because the number of possible matchings 
grow with the number $n$ of nodes resulting in an increasing  variance.

This result shows that the sampling error does not averages out. Hence, it is not meaningful to investigate sampled overlap distributions $P(q)$ with samples directly drawn from the perturbation process.

\section{Conclusion}
\label{sec:Conclusion}

We have studied the behavior of the energy landscape of matchings 
 by using \afz{edge flips} to perturb the optimal solution of the original
graphs. Our analyses have shown that for suitable ensembles  there exist quasi optimal states with differ from the ground state by an extensive number of edges. Hence, a complex energy landscape can not be excluded here.
These ensembles include ring graphs with a constant or logarithmic growing number of additional edges and constant edge weights with uniform noise. In these cases, the extrapolated relative distance $d_\infty$ between the matchings stays finite.
For the other observed ensembles, ring graphs with linear growing number of extra edges and for \ER\ graphs, the results show no evidence for a complex energy landscape.

It was shown that samples taken directly from the perturbation method are biased. This hindered further more detailed investigations,
like obtaining the distribution of overlaps between many optimum matchings.
 Hence, it is unknown if the quasi optimal states are relevant in the thermodynamic limit. If this was not the case, the behavior of the energy landscape 
observed here resembles ``weakly broken replica symmetry'' 
\cite{Parisi_1990_weakRSB}. It should be noted that, in general,
approaches based on perturbing given optimum solutions are only able
to show whether there exist other solutions which differ by $O(n)$
variables, but not whether they are relevant in the thermodynamic limit.
Still, if these very different other quasi-optimal solutions do not exist, 
it is clear that the energy landscape is simple.

To set up further studies, it should be noted that for ring graphs with 
a constant or logarithmically growing number $m^+(n)$ of additional
 edges, a finite value of $d_\infty > 0$ was observed. But if $m^+(n)$ 
grows linearly with $n$, we found $d_\infty = 0$. Consequently, there exist a transition between choices for $m^+(n)$ where $d_\infty$ remains finite and where it vanishes. To analyses this further, the number of extra edges could
 be modeled by a power law $m^+ \in O(n^\lambda)$. Then, it can be evaluated 
if a critical value $\lambda_c$ for the exponent exists, 
where a sudden transition between $d_\infty > 0$ and $d_\infty = 0$ takes place.

Also, it would be interesting to find a method to efficiently draw unbiased samples. One could use sampling methods
which  are numerically more demanding, like Monte Carlo
Markov-chain simulations \cite{newman1999} or exhaustive enumerations.
With present algorithms this would restrict the accessible system sizes
considerable.
 On the other hand, it could be possible to correct the sampling bias
afterwards using the \emph{ballistic-search} approach \cite{ballistic2000}, 
which
has been applied successfully to correct the sampling bias observed for 
evolutionary algorithms when applied to spin glasses \cite{excited2002}.
This is algorithmically quite demanding, but certainly feasible in 
future projects. 
By applying such approaches, the sampled overlap distributions $P(q)$ 
can then be investigated to get more inside into the energy landscape of the 
ensembles proposed here. In this way, 
it would also possible to check whether the quasi-optimal very different
solutions are relevant in the thermodynamic limit.

Beside matching and directed polymers, other P problems may also show indications of a complex energy landscape for suitable ensembles. The methods described in this or other works  
\cite{ricci-tersenghi2001,mezard2003,Hartmann_2022_polymers,is_rsb2023} can 
be adjusted to these problems to studied them in a similar fashion.

\ack  
The simulations were partially performed at the HPC cluster CARL, located at the
University of Oldenburg (Germany) and funded by the DFG through its Major
Research Instrumentation Program (INST 184/157-1 FUGG) and the Ministry of
Science and Culture (MWK) of the Lower Saxony State.

\bibliography{references}

\begin{thebibliography}{10}
\bibitem{matching_is_numberP}
{In \cite{Valiant_1979_numberP} is was shown that perfect matchings on
  bipartite graphs are \#P-complete. Each graph has a bipartite subgraph and
  this subgraph has at least as many matchings as perfect matchings. Hence, it
  follows that also matching is \#P-complete.}
\end{thebibliography}

\end{document}